\documentclass[%
    reprint,
    superscriptaddress,
    amsmath,
    amssymb,
    aps,
  floatfix,
]{revtex4-2} 

\usepackage{graphicx}
\usepackage{dcolumn}
\usepackage{bm}
\usepackage{units}
\usepackage{xcolor}
\usepackage{soul}
\usepackage{cases}

\usepackage[normalem]{ulem}
\usepackage{empheq}

\newcommand{\elD}[2]{\dfrac{\delta #1}{\delta #2}}

\newcommand{\x}{\bm{x}}
\newcommand{\y}{\bm{y}}

\newcommand{\X}{\bm{X}}

\newcommand{\e}{\hat {\bm{e}}}
\newcommand{\q}{\bm{q}}
\newcommand{\vi}{\bm{v}}
\newcommand{\n}{\hat{\bm{n}}}

\newcommand{\Fr}{\mathcal{F}} 
\newcommand{\effM}{\mathcal{M}}
\newcommand{\brp}{\bar \phi}

\newcommand{\bt}{\bar\theta}
\newcommand{\Dbp}{\Delta\bar \phi}
\newcommand{\bfeta}{\boldsymbol{\eta}}
\newcommand{\Lam}{\bm{\Lambda}}
\newcommand{\sX}{\bm{\mathcal{X}}}

\newcommand{\lt}{\left}
\newcommand{\rt}{\right}

\DeclareMathAlphabet{\pazocal}{OMS}{zplm}{m}{n}

\newcommand{\shortarrow}[1][4pt]{\mathrel{%
   \hbox{\rule[\dimexpr\fontdimen22\textfont2-.2pt\relax]{#1}{.4pt}}%
   \mkern-4mu\hbox{\usefont{U}{lasy}{m}{n}\symbol{41}}}}
\newcommand{\ch}[2]{\mathcal{C}_{#1\shortarrow #2}}
\newcommand{\chot}{\ch{1}{2}}
\newcommand{\chto}{\ch{2}{1}}

\usepackage[ISO]{diffcoeff}
\usepackage{bm}
\usepackage{hyperref}

\newcommand{\I}{(\mathrm{I})}
\newcommand{\II}{(\mathrm{II})}
\newcommand{\III}{(\mathrm{III})}
\newcommand{\dd}{\mathrm{d}}

\begin{document}

\title{Non-reciprocal interactions between condensates in chemically active mixtures}

\author{Jacopo Romano}
\affiliation{Max Planck Institute for Dynamics and Self-Organization (MPI-DS), 37077 G\"ottingen, Germany}
\affiliation{SISSA --- International School for Advanced Studies and INFN, via Bonomea 265, 34136 Trieste, Italy}

\author{Martin Kjøllesdal Johnsrud}
\affiliation{Max Planck Institute for Dynamics and Self-Organization (MPI-DS), 37077 G\"ottingen, Germany}

\author{Beno\^it Mahault}
\affiliation{Max Planck Institute for Dynamics and Self-Organization (MPI-DS), 37077 G\"ottingen, Germany}

\author{Ramin Golestanian}
\affiliation{Max Planck Institute for Dynamics and Self-Organization (MPI-DS), 37077 G\"ottingen, Germany}
\affiliation{Rudolf Peierls Centre for Theoretical Physics, University of Oxford, Oxford OX1 3PU, United Kingdom}

\date{\today}

\begin{abstract}
We study the behaviour of catalytically active droplets in multi-component conserved mixtures affected by noise. Working in the thin interface limit, we analytically determine the state diagram of the system, characterized by multiple dynamical regimes, and verify our findings using numerical simulations. In particular, we show the emergence of a non-reciprocal, chemically-mediated interaction between the droplets, which leads to the formation of (meta-)stable clusters of droplets of different species. We find that the clusters can display self-propulsion in a large part of the parameter space, including regions where the non-reciprocal interactions between the droplets are purely attractive. This surprising feature arises from the non-local nature of the chemical interactions, and points to locality violations as a general mechanism for energy dissipation and emergence of out-of-equilibrium steady states in active matter. 

\end{abstract}

\maketitle

The distinguishing feature of an active system is the continuous energy consumption of its microscopic constituents, which drives the system out of equilibrium leading to complex and novel interactions. In multiple experimental systems---both biological and artificial---such interactions are due to the production and consumption of chemicals, which in turn affect the dynamics of the species constituting the system \cite{golestanianPhoreticActiveMatter2022,Stark2018,AgudoCanalejo2018a,popescu2016self,Weber2019}. Examples include large collections of catalytically active colloids 
\cite{Golestanian2005,golestanianPhoreticActiveMatter2022,howse2007, paxton2004catalytic}, bacterial colonies and collections of cells  \cite{patteson2015running,tallieur2008mips,cates2015motility,Gelimson2015,Mahdisoltani2021a}, as well as metabolically active condensates \cite{brangwynne2011active,Testa2021,jambon2024phase,Dindo2025,guo2024liquid,demarchi2023enzyme,Kokkoorakunnel2024}, and suspensions of enzymes \cite{OuazanReboul2023b,OuazanReboul2023,OuazanReboul2023a}. Chemical activity of enzymes manifests in a variety of non-equilibrium behaviour including growth and division of droplets \cite{Zwicker2015ripening,zwicker2017growth,Golestanian2017,donau2023trends,modiDesigningNegativeFeedback2023,modiTransientPHChanges2025}. Such mixtures are naturally present in the cytoplasm of cells and it is believed that their activity is crucial in allowing them to perform their biological functions \cite{brangwynne2011active,shin2017liquid}. 

\begin{figure}[!b]
    \centering
    \includegraphics[width=\linewidth]{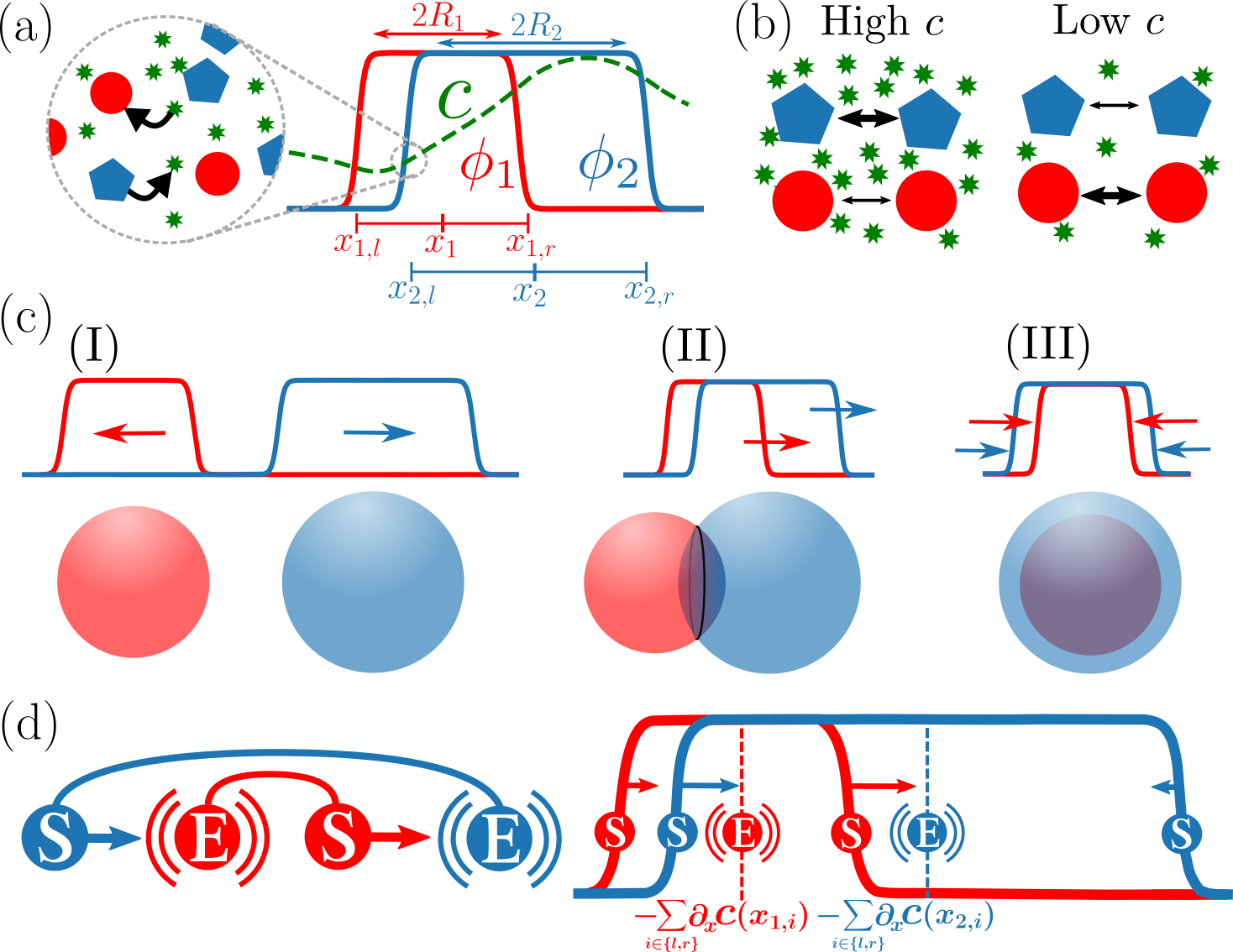}
    \caption{
    (a) The species gather into droplets due to their interaction. Spatial variations in the densities of the species ($\phi_1$ and $\phi_2$) create a spatially varying chemical field ($c$, in green) as they produce and consume the chemical. 
    (b) The chemical affects the intra-species interaction and mediates effective inter-species interactions.
    (c) Configurations of droplets: (I) disjoint, (II) partially overlapping and (III) fully overlapping.
    (d) Attractive interactions give rise to chasing-behaviour.
    The centre of the droplet is an emitter (E) of a chemical signal and the interfaces are sensors (S).
     An example where purely attractive interactions lead to chasing (left) and its implementation with two droplets (right).
    }
    \label{fig:possdroplets}
\end{figure}

When multiple chemically active species are present in a mixture, the effective interactions mediated by the chemical often break action-reaction symmetry \cite{Soto2014,Saha2020,you2020nonreciprocity,Ivlev2015brokenthird,Uchida2010a,fruchart2021non}. This has been studied in droplets \cite{meredith2020predator}, mixed bacterial colonies \cite{duanyu2023patterns,natan2022mixed,dinelli2023non} and catalytic colloids \cite{Soto2015,AgudoCanalejo2019,OuazanReboul2021,Tucci2024}, showing the formation of self-propelled clusters. The formation of multi-component, multi-phase mixtures has been observed also in vivo in embryo cells \cite{feric2016multiphase}. The physical mechanisms describing how the chemicals affect the species are not fully understood, and are likely system-specific, especially for phase separating biopolymers \cite{donau2023trends}. It has been argued that the enzymes in these mixtures display an effective chemotactic behaviour \cite{AgudoCanalejo2018}, leading to a mass current of the species proportional to the chemical gradients \cite{WalterPRL2017,demarchi2023enzyme,Cotton2022,Goychuk2024sharp,zhao2023chemotactic}.
Moreover, it has been shown in multiple experiments on phase separating droplets \cite{Testa2021,jambon2024phase,Dindo2025,banani2017biomolecular,kim2023rna,brangwynne2009germline} that the chemicals affect the species by altering the the inter-species interactions, which determine the phase separation, effectively changing the critical point of the mixture.
These observations cannot be fully accounted for by chemotactic currents, and suggest that the free energy of the mixture depends on the local chemical concentrations, as illustrated in Figs.~\ref{fig:possdroplets}(a) and \ref{fig:possdroplets}(b).

In this Letter, we explore the dynamics of a multi-component mixture in the presence of a chemical which affects its material properties and is in turn produced and consumed by the components of the mixture.
We elucidate how such dependencies lead the individual droplets to respond to chemical gradients, reminiscent of phoretic motion. We then demonstrate how this property gives rise to effective non-reciprocal interactions between droplets in a way that can be controlled via material parameters and geometric characteristics.
We find the emergence of (meta-)stable clusters of droplets of different species, displaying different mechanisms of self-propulsion. The first mechanism is chasing of droplets due to non-reciprocal, attractive-repulsive interactions, similarly to what is observed in \cite{Soto2014,AgudoCanalejo2019,duanyu2023patterns,dinelli2023non}. The second mechanism is non-locality-induced chasing of mutually attractive  droplets, defying the expectation that self-propulsion requires attractive-repulsive interactions. Finally, the third mechanism is due to the interplay between non-reciprocal couplings and fluctuations in deterministically static configurations.  

\textit{Theoretical framework}---To describe the system's dynamics, we model the evolution of $N$ density fields for $\phi_a(\bm r,t)$, $a\in\{1, 2, \cdots, N\}$, which are governed by a free energy $\Fr[\{\phi_a\},c]$ that depends on a chemical field $c$ that the species produce or consume, e.g.~through dependence of the material properties on the chemical. We neglect advective transport, which is appropriate for low-dimensional or sufficiently small systems \cite{romano2025dynamics,SuppMat} and qualitatively justified in regimes exhibiting ripening phenomenology~\cite{brangwynne2009germline,berry2018physical,tateno2026diffusive}. We thus consider a generalization of model B in $d$ dimensions \cite{HalperinHohenberg1977}, as follows 
\begin{subnumcases}{\label{eq:eom}}
    \label{eq:LRNRCH}
    \partial_t\phi_a=\bm \nabla\cdot\left[M_a\bm \nabla\elD{\Fr[\{\phi_b\},c]}{\phi_a}\right]+\sqrt{2 M_a T}\bm \nabla\cdot\Lam_a,\\
    \label{eq:chemical}
    -D_c{\bm \nabla}^2 c+\gamma c=\sum_a r_a\phi_a,
\end{subnumcases}
where $M_a$'s are the corresponding mobilities, $\Lam_a$'s are Gaussian stochastic currents with zero mean and correlations $\langle \Lambda_{a,i}(\bm r,t)\Lambda_{b,j}(\bm r',t')\rangle = \delta_{ij} \delta_{ab} \delta^d(\bm r-\bm r') \delta(t-t')$ \cite{kubo1966fluctuation}, and $T$ is the temperature. 
In Eq.~\eqref{eq:chemical}, $r_a$'s are the production (consumption) rates of the chemical by the two species, $D_c$ is the diffusion coefficient of the chemical, and $\gamma \equiv D_c \kappa^2$ is its decay rate, which introduces a decay length scale $\kappa^{-1}$. The dynamics of $c$ is assumed faster than that of the two species; in this limit, temporal evolution and fluctuations of $c$ can be neglected~\cite{Mahdisoltani2021a,huang2021microscopic}, leading to~\eqref{eq:chemical}.
For simplicity, we also assume that the free energy can be decomposed into independent parts: $\Fr[\{\phi_b\},c]=\sum_a \Fr_a[\phi_a,c]$. In Appendix \ref{appendix 1}, we show in that our results are not fundamentally modified by the presence of (weak) inter-species interactions. We then consider: $\Fr_a=\int d^d\bm r  \left[\frac{1}{2} K_a(\phi_a,c) ({\bm \nabla}\phi_a)^2+U_a(\phi_a,c) \right]$ where the $U_a$ terms are double-well potentials (see Appendix \ref{appendix 2}) promoting phase separation in both species leading to the formation of droplets of the dense phase coexisting with a dilute background.

In the sharp interface limit \cite{Bray2002review}, the densities in and out of the droplets are given by the two minima of the field potentials $U_a(\phi_a)$, which we denote by $\brp_{a,+}$ for the dense phases and $\brp_{a,-}$ for the dilute phases, leading to density contrasts of $\Delta \brp_{a}=\brp_{a,+}-\brp_{a,-}$ across the droplet interfaces. The dynamics of the droplets is then be controlled by $c$-dependent surface tensions $\sigma_a(c)$, which can be calculated in terms of $U_a$ and $K_a$, as well as $\bar \phi_{a,\pm}(c)$ \cite{bray1993kinetics, romano2025dynamics}. We consider the linear response regime, in which $\sigma_a(c)=\sigma_{a}(0)+\frac{\partial\sigma_a}{\partial c} c$ with the derivative evaluated at vanishing concentration, and neglect the $c$-dependence of $\bar \phi_{a,\pm}$ (see SM \cite{SuppMat}).

Our study reveals that the coupling constants
\begin{equation}
    \ch{a}{b}\equiv -r_{b} \cdot \frac{M_a}{\Delta \brp_{a}^2} \frac{\partial\sigma_a}{\partial c},
\end{equation}
specifies the strength of the interaction experienced by $a$ toward $b$, with the convention that for positive $\ch{a}{b}$ species $a$ is attracted to species $b$. This is because a droplet of species $a$ climbs (or descends) chemical gradients at a rate proportional to $-\frac{M_a}{\Delta \brp_{a}^2} \frac{\partial\sigma_a}{\partial c}$, while it senses a chemical gradient proportional to the production rate $r_b$ of the other droplet [see Eq.~\eqref{eq:chemical}]. As in general we have $\ch{a}{b}\neq\ch{b}{a}$, the interactions are non-reciprocal, similar to what is observed in phoretic active systems \cite{Soto2014,AgudoCanalejo2019}. These droplets, however, present remarkable additional features in comparison with phoretic colloids, as their effective non-reciprocal interactions can be controlled by tuning their geometric features and the screening length, which is afforded by the emergent non-trivial size dependence of the effective mobilities (see below). Moreover, the droplets can penetrate each other and lead to a richer set of configurations for active complexes that form due to non-reciprocal interactions (see Fig.~\ref{fig:possdroplets}). We note that the non-reciprocity, which manifests itself both in strength and sense of the interactions, cannot be simply represented by an anti-symmetric matrix of coupling constants due to the non-trivial dependence on the droplet sizes and screening length.

To elucidate this phenomenology, we focus on the long-time behaviour of the system, which is characterized by the presence of spherical droplets of radii $R_a$ (one for each species), with volume $\pazocal{V}_a=V_d R_a^d$ and surface area $\pazocal{S}_a=d V_d R_a^{d-1}=d \pazocal{V}_a/R_a$, where $V_d=\pi^{d/2}/\Gamma(\frac{d}{2}+1)$.
We can use the dynamical equation for the concentration fields \eqref{eq:eom} to extract effective stochastic equations of motion for the centre of mass of the droplet of species $a$, which we denote by $\x_a$ (see Fig. \ref{fig:possdroplets}), as follows \cite{romano2025dynamics}
\begin{equation}
    \label{eq:x-dyn-eq}
    \dot \x_a=-\effM_a {\cal S}_a \frac{\partial\sigma_a}{\partial c}  {\bm \nabla} {\bar c}_{{\rm s},a}(\x_a) +\sqrt{2\effM_a T}\,\bfeta_a(t), 
\end{equation}
where ${\bar c}_{{\rm s},a}(\x_a)\equiv \langle c\rangle_{{S}_a}=\frac{1}{d {V}_d}\int d\n_a c(\x_a+R_a\n_a)$ with $\n_a$ representing the radial unit vector describing the points on the boundary $S_a$ of droplet $a$, $\effM_a=M_a d/(\pazocal{V}_a  \Delta \brp_{a}^2)$ is an effective mobility coefficient, and $\bfeta_a$'s are Gaussian white noise terms with zero mean and correlations $\langle \eta_{a,i}(t)\eta_{b,j}(t')\rangle = \delta_{ij} \delta_{ab}  \delta(t-t')$. We note that Eq. \eqref{eq:x-dyn-eq} is independent of the details of the free energy, and can be regarded as a generalized (diffusio-)phoretic transport mechanism resulting from interfacial stresses \cite{anderson1989}. Note that Eq.~\eqref{eq:x-dyn-eq} describes the dynamics of droplets even when they overlap. Within the thin interface approximation, it is possible to derive exact expressions for ${\bar c}_{{\rm s},a}(\x_a)$ in different dimensions (see Appendix \ref{app:d-dim-phoresis}).

When the droplets do not overlap, we obtain a closed-form expression for the surface-averaged concentration and use it to express the drift velocity as follows 
\begin{align}
   \hskip-1.53mm \dot \x_a=- \frac{d^2 M_a}{\Delta \brp_{a}^2} \frac{\partial\sigma_a}{\partial c}
    \frac{{\cal I}_d\left(\kappa R_a\right)}{R_a} {\bm \nabla} c(\x_a) +\! \sqrt{2\effM_a T}\bfeta_a(t),\label{eq:phoresis-result-d}
\end{align}
where ${\cal I}_d(b)$ is defined in Appendix \ref{app:d-dim-phoresis}, which details the derivation.
To demonstrate the emergence of non-reciprocal droplet interactions and the resulting phenomenology, we focus on the two-body dynamics of two droplets of different species. We define the relative position of the two droplets as $\x=\x_2-\x_1$ and their midpoint as $\X=\frac{1}{2}(\x_1+\x_2)$. The dynamics of these coordinates will take the general form
\begin{subnumcases}{\label{eq:dynLangevin}}
    \label{eq:dynpotential}
    \dot \x=-\effM\bm \nabla_{\x} W_{-}(\x)+\sqrt{2 \effM T}\,\bfeta_x(t),\\
    \label{eq:dynCoM}
    \dot \X=\vi(\x)+\sqrt{\effM T/2}\,\bfeta_X(t),
\end{subnumcases}
where $\bfeta_x(t)$ and $\bfeta_X(t)$ are two independent Gaussian white noises with zero mean and unit variance, $\effM=\effM_1+\effM_2$, and $\vi(\x)=-\frac{1}{2}\effM\bm\nabla_{\x} W_{+}(\x)$, with the potentials $W_{\pm}(\x)=\frac{\effM_2}{\effM} {\cal S}_2 \frac{\partial\sigma_2}{\partial c} {\bar c}_{{\rm s},2}(\x) \pm \frac{\effM_1}{\effM} {\cal S}_1 \frac{\partial\sigma_1}{\partial c}  {\bar c}_{{\rm s},1}(\x)$ depending only on $\x$ due to translational invariance of the system. 
We start with $d=1$, where a droplet of species $a$ is characterized by the positions $x_{a,l}=x_a-R_a$ and $x_{a,r}=x_a+R_a$ of its left and right interfaces, respectively. Consequently, we can obtain a simple expression for the surface-averaged concentration as ${\bar c}_{{\rm s},a}(x_a)=[c(x_a-R_a)+c(x_a+R_a)]/2$.
Note, however, that this choice is made for simplicity of the presentation, and that analytical results can be obtain in all dimensions, as shown in Appendix \ref{app:d-dim-phoresis}.
From here onward, We set $\brp_{1,+}=\brp_{2,+}=1$, $\brp_{1,-}=\brp_{2,-}=0$, $D_c=1$, and $\gamma=1$, and hence express all length scales in terms of the decay length of the chemical $\kappa^{-1}=\sqrt{{D_c}/{\gamma}}=1$ \cite{SuppMat}. Moreover, we assume $R_2>R_1$ without loss of generality. 
To close the equations, we need to solve for the concentration field using Eq.~\eqref{eq:chemical}, which yields $
\partial_x c(x)=\frac{1}{2}\sum_{a}r_a\left(e^{-|x-x_a+R_a|}-e^{-|x-x_a-R_a|}\right)$.

Using this expression, we obtain explicit expressions in the form of  Eq.~\eqref{eq:dynLangevin}, with the potentials $W_{\pm}(x)$ given in Appendix \ref{app:1D}. As anticipated, the expressions for $W_{\pm}(x)$ depend on the material parameters only through the coupling constants $\ch{a}{b}$, highlighting the non-reciprocal nature of the interactions.

\begin{figure}[!tb]
\centering\includegraphics[width=\columnwidth]{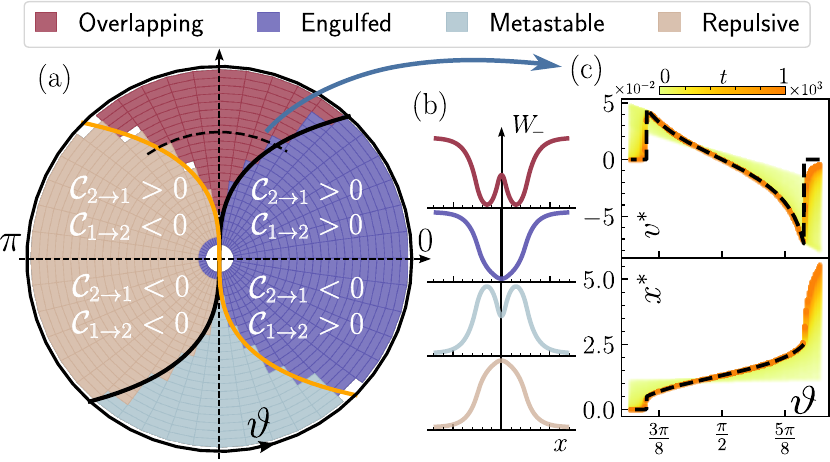}
    \caption{
        (a) State diagram from simulations of Eq.~\eqref{eq:eom}, presented a radial plot with radius $\mathcal{R} = R_1 / R_2$ and angle $\vartheta = \arg(\chot+\mathrm{i}\,\chto)$. The black and orange lines mark the boundaries of Eq.~\eqref{eq:stabingl} and Eq.~\eqref{eq:stabrep} respectively. 
    (b) Effective potentials for $W_{-}(\bm x)$ in the various states.
    (c) Steady-state velocity and relative position calculated along the dashed line in (a) (dashed black), compared with direct simulations. Colour indicates time, with the steady-state shown in orange.
    } 
    \label{fig:phase}
\end{figure}

\textit{Deterministic dynamics}---We can now study the behaviour of the system as a function of chasing rates and droplet sizes. In Fig.~\ref{fig:phase}(a), we present a classification of the different states of the dynamics by using the dimensionless parameters $\mathcal{R} \equiv R_1 / R_2$ as radius and $\vartheta \equiv \arg(\chot+\mathrm{i}\,\chto)$ as angle (see Appendix \ref{appendix 2} for the details of the rules used in the classification).

Starting our analysis with no noise, we note that Eq.~\eqref{eq:dynLangevin} always admits $x=0$ as a stationary point, corresponding to the symmetric configuration with the small droplet encapsulated inside the large one, i.e.  configuration (III) of Fig.~\ref{fig:possdroplets}(c). This fixed point is stable when
\begin{equation}
    \label{eq:stabingl}
    \chto 
    <
    \frac{R_2}{R_1}
    \frac{1}{\tanh R_1}\,
    \chot,
\end{equation}
which means that if Eq.~\eqref{eq:stabingl} is satisfied, droplets will reach a symmetric configuration of complete engulfment when starting from asymmetric initial conditions [see Fig.~\ref{fig:phase}(a), black line]. For sufficiently large initial distances [droplets in configuration (I) of Fig.~\ref{fig:possdroplets}(c)], the potential decreases with $|x|$ when
\begin{equation}
    \label{eq:stabrep}
     \chto < - \frac{R_2}{R_1} \dfrac{\tanh R_2}{\tanh R_1 } \,\chot.
\end{equation}
In this case, two well-separated droplets experience an effective repulsion (and their distance increases indefinitely). Note that Eqs.~\eqref{eq:stabingl} and~\eqref{eq:stabrep}, which can be satisfied at the same time, become increasingly symmetric around the $\chot=0$ line [vertical axis in Fig.~\ref{fig:phase}(a)] for large values of $R_1$ and $R_2$. Eqs.~\eqref{eq:stabingl} and~\eqref{eq:stabrep} can be satisfied at the same time. In this case, $W_{-}(x)$ has a metastable minimum at $x=0$, while disjoint droplets repel each other [see Fig.~\ref{fig:phase}(b)]. Finally, when both Eq.~\eqref{eq:stabingl} and $\eqref{eq:stabrep}$ are violated, $W_{-}(x)$ becomes a double-well potential with two minima [see Fig.~\ref{fig:phase}(b)], representing configuration (II) of Fig.~\ref{fig:possdroplets}(c). In this case, the fixed point at $x=0$ is unstable, while two stable fixed points appear at $\pm x^*$, where $x^* = \frac{1}{2}\ln \left[1+\left(e^{2R_2}-e^{2R_1}\right)\frac{\chto R_1-\chot R_2}{\chto R_1+\chot R_2}\right]$.
This corresponds to the partially overlapping configuration (II) of Fig.~\ref{fig:possdroplets}(c), with the smaller droplet being on the left of the larger one for $x >0$.
The steady-state velocity $v^*=v(x^*)$ of the midpoint of the cluster is then found $v^*=\frac{\sinh\left(R_2-R_1\right)\, \chot \,\chto}{(\chto R_1+\chot R_2)}e^{-x^*}$ [see Eq.~\eqref{eq:dynCoM}].
These predictions are verified by simulating Eq.~\eqref{eq:eom} at negligible noise ($T=10^{-5}$) to check if the final state is in a disjoined, overlapping, or engulfed configuration [respectively, in configurations (I), (II), or (III) of Fig. \ref{fig:possdroplets}(c)], and it is indeed observed that the shape of the effective potential $W_{-}$ determines the behaviour; see Figs.~\ref{fig:phase}(a), ~\ref{fig:phase}(b), and \ref{fig:phase}(c).

It is remarkable that self-propelling clusters are observed in the region of fully attractive, albeit non-reciprocal, droplet interactions where $\chot>0$ and $\chto>0$. This may be surprising, as self-propulsion in systems with short-range interactions emerges when the chasing rates have opposite signs. However, since droplets are extended objects, additional opportunities to realize conditions for effective non-reciprocal interactions arise from the fact that the sources and sinks of chemicals are in the bulks of the droplets while the force-sensing operates at their interfaces. The chasing rates determine how the interface of each droplet interacts with the centre of the other. This implies that configurations in which the leftmost interface of droplet $2$ lies between the leftmost interface and centre of mass of droplet $1$ can lead to a net repulsive effect on droplet $1$, despite all the forces being attractive from the point of view of the respective interfaces; see Fig.~\ref{fig:possdroplets}(d). Note that this effect is only possible in active systems. 

As a consequence of the change in self-propulsion mechanism when $\chot$ changes sign (corresponding to $\vartheta=\frac{\pi}{2}$) the cluster inverts its velocity. While for $\chot<0$ droplet $1$ on the left is repelled by droplet $2$ that is chasing it, for $\chot>0$ droplet $1$ chases droplet $2$ due to the effective non-reciprocity described above. This is shown in Fig.~\ref{fig:phase}(c), where we present the velocities and relative positions for a set of simulations, within, and just outside, the overlapping regime. In both cases, the solutions relax over time towards the analytical predictions. 

\begin{figure}[!tb]
\centering\includegraphics[width=\columnwidth]{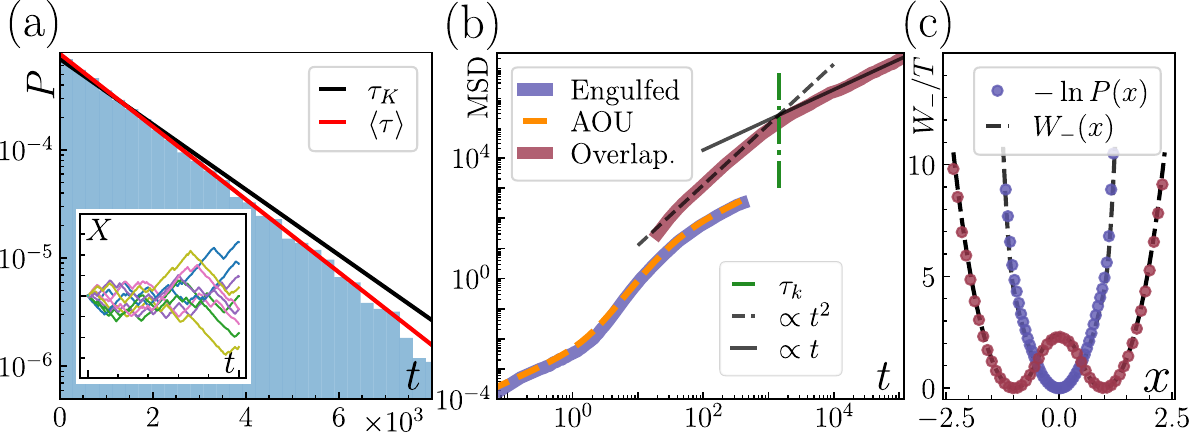}
    \caption{
    (a) 
    Jumping times for the relative position in the overlapping regime at $T=2\times10^{-2}$, following a Poisson distribution 
    $\langle\tau\rangle = 1.29 \times 10^{3}$, compared with $\tau_K = 1.44\times 10^3$. Inset: Centre position against time for 10 droplet pairs.
    (b) MSD of the cluster from simulations of Eq.~\eqref{eq:eom} in the overlapping (red) and engulfed (blue) regimes. 
    The latter follows an AOU process with coefficients obtained from Eq.~\eqref{eq:dynLangevin} (dashed orange).
    (c) Steady-state probability distribution for $x$. Points show simulations of overlapping (red) and attractive (blue) regimes, compared with Eq.~\eqref{eq:dynLangevin} (dashed lines).
    }
    \label{fig:stochastic}
\end{figure}

\textit{Stochastic dynamics}---We now study the finite temperature case, focusing on the attractive and overlapping regimes. In the overlapping regime of Fig.~\ref{fig:phase}, Eq.~\eqref{eq:dynpotential} describes a Langevin dynamics in a double-well potential. At sufficiently low temperatures, the dynamics of the relative position is confined in the close neighbourhood of the two minima $\pm x^*$, while rare trajectories exhibit barrier-crossing, going from one minimum to the other. The mean first-passage time can be calculated from Kramers' law as
$\tau_K=\frac{2\pi \exp\left\{\left[W_{-}(0)-W_{-}(x^*)\right]/T\right\}}{\sqrt{|W_{-}''(0)W_{-}''(x^*)|}}$ \cite{Kramers1940,Eyring1935}. Each time the relative position passes from one minimum to the other, the cluster midpoint $X$ inverts its direction of motion. This results in an {\em emergent} run-and-tumble dynamics~\cite{Bennett2013,Soto2015}, which can be modelled as $\dot X=v^* e(t)+\sqrt{\effM T/2}\,\eta_X(t)$, where $e(t)=\pm 1$ is a random Poisson process with rate $1/\tau_K$ characterizing the tumbling motion~\cite{Demaerel2018tumble}. The probability distributions for the stochastic jump times and relative positions obtained from our simulations of Eq.~\eqref{eq:eom} show good agreement with the prediction from Kramers theory [see Fig.~\ref{fig:stochastic}(a)] and the Boltzmann distribution associated with the potential $W_{-}$ [see Fig.~\ref{fig:stochastic}(c)]. The run-and-tumble dynamics of the midpoint is then characterized by a transition of the mean-squared-displacement (MSD) $\langle X(t)^2 \rangle$ from ballistic to diffusive behaviour at $\tau_K$, which is is consistent with the behaviour observed from Eq.~\eqref{eq:eom}, as shown in Fig.~\ref{fig:stochastic}(b).

In the engulfed state at zero temperature, the steady-state is a static configuration with the smaller droplet symmetrically placed inside of the larger one, leaving no sign of the chemical activity in the droplet dynamics at large times. However, when temperature is increased, the relative position starts to fluctuate around $x=0$, which in turn affects the dynamics of the midpoint. At small temperatures we can expand Eq.~\eqref{eq:dynCoM} around $x=0$, obtaining the equations of motion of an active Ornstein-Uhlenbeck process (AOU) \cite{nguyen2021active,martinStatisticalMechanicsActive2021}. We thus retrieve a signature of the chemical activity of the system in the MSD of the midpoint, and quantify it analytically
and via numerical simulations of Eq.~\eqref{eq:eom} [see Figs.~\ref{fig:stochastic}(b) and \ref{fig:stochastic}(c)]. We find that active fluctuations enhance the diffusivity of the engulfed clusters at large times, broadening the range of parameters for which the droplet dimer exhibits self-propulsion. Such active fluctuations can provide the possibility to control the self-organization of these compound active particles \cite{Mahault2023}.

\textit{Higher dimensions}---We now briefly examine the role of dimensionality to probe the robustness of our findings. Due to rotational symmetry, $W_{-}$ is a radial function, which can be a attractive, repulsive, or in the form of (inverted) Mexican hat potential, depending on the radii and chasing rates. When the potential forms a Mexican hat, its degenerate minimum is a circle or sphere of radius $x^*$. When the potential is strong as compared to the noise, the dynamics is confined to this minimum, such that the relative position satisfies $\x(t)\simeq x^* \e(t)$, where $\e(t)$ is a unit vector, which evolves according to $\dot\e(t)=\frac{1}{x^*}\sqrt{2 \effM T}\,\e(t)\times[\e(t)\times\bfeta_x(t)]$. 
As before, the dynamics of the midpoint reads $\dot \X=|\vi(x^*)|\e(t)$ (ignoring its own noise term). We thus find that the self-propelled cluster behaves effectively like an active Brownian particle \cite{romanczuk2012active,howse2007} with angular diffusion coefficient $T \effM/(x^*)^2$. 

\begin{figure}[t]
\centering\includegraphics[width=\linewidth]{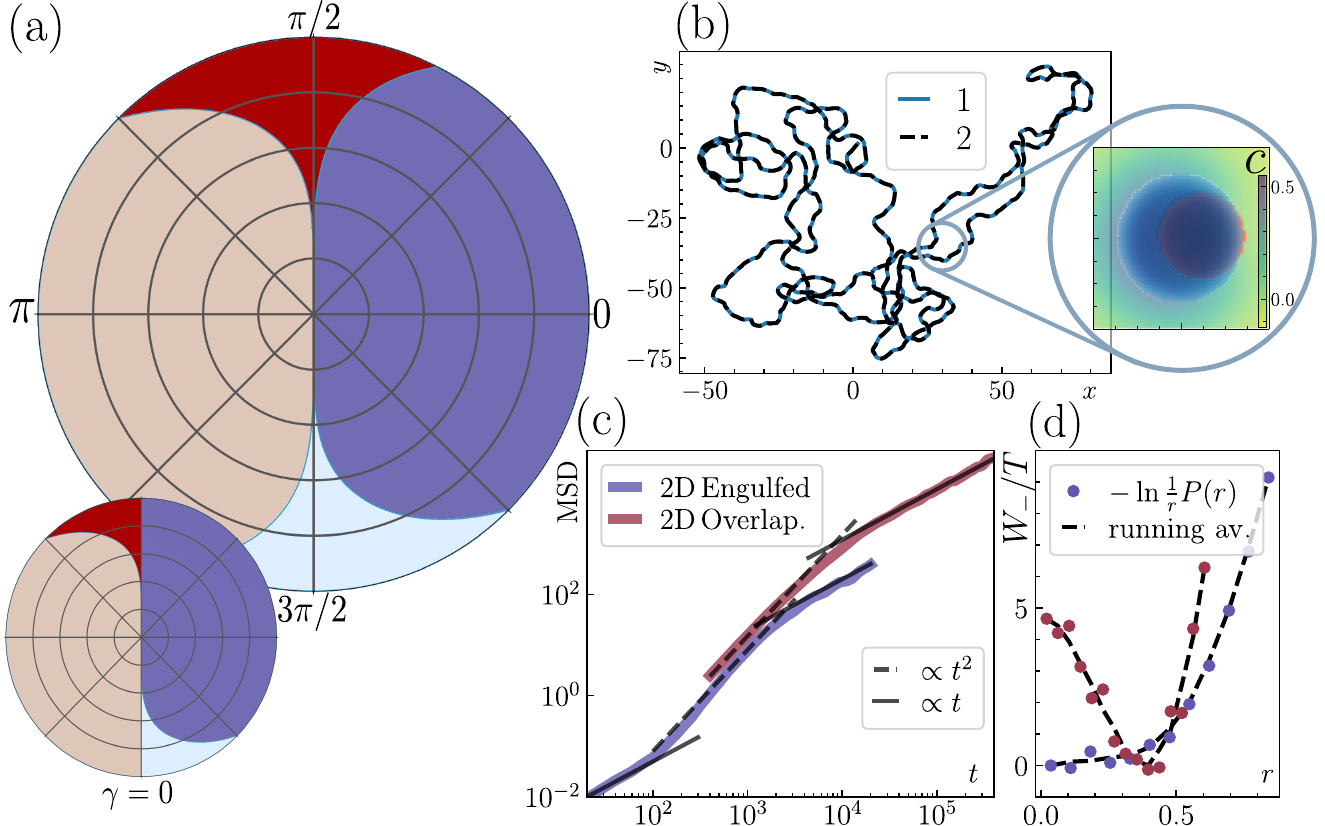}
    \caption{
    (a) State diagram at finite (large disk) and vanishing (small disk) decay rate $\gamma$ in $d=3$, colours as in Fig.~\ref{fig:phase}.
    (b) Simulations of Eq.~\eqref{eq:eom} in $d=2$ at $T = 10^{-4}$, droplet 1 (2) in blue (black).
    Inset: Snapshot of the densities $\phi_a$ and the chemical field $c$. (c) MSD for droplets in the attractive (blue) and overlapping (red) configurations. (d) The effective potential for $\x$, inferred from the probability distribution of the relative position.
    Dashed lines are running averages.
    }
    \label{fig:pasedia3ddecay}
\end{figure}

We analytically calculate $W_{-}$ in $d=3$ and use it to build the state diagram (see Appendix \ref{app:d-dim-phoresis} and Ref. \cite{SuppMat}), which results qualitatively similar to the 1D case. At finite decay rate $\gamma$ [Fig.~\ref{fig:pasedia3ddecay}(a)], non-locality allows for the formation of self-propelling clusters of mutually attractive droplets, highlighting the general nature of this mechanism and its relevance to the dynamics of multi-component mixtures. Working in 3D allows us to study Eq.~\eqref{eq:eom} in the case of $\gamma=0$, which is characterized by the lack of overlapping clusters at $\chot, \chto > 0$ [Fig.~\ref{fig:pasedia3ddecay}(a)].

While the same procedure could be carried over also in $d=2$ (see Appendix \ref{app:d-dim-phoresis}), we choose to take a complementary approach and present a numerical simulation of Eq.~\eqref{eq:eom} in this case. Again, in 2D we find overlapping, self-propelling clusters of droplets with both positive chasing rates [Fig.~\ref{fig:pasedia3ddecay}(b)], as well as attractive clusters with AOU behaviour. The MSD versus time plot [Fig.~\ref{fig:pasedia3ddecay}(c)] shows a transition from ballistic to diffusive motion at a time consistent with our analysis. We also measure the distribution of the separation of the droplets, which in 2D is related to the effective potential as $W_-(\x)\propto - \ln \frac{1}{x}P(x)$, to estimate the effective potential in the engulfed and overlapping regimes [Fig.~\ref{fig:pasedia3ddecay}(d)].

\emph{Discussion}---We have investigated the dynamics of catalytically active multi-component mixtures. Our study indicates that droplet motion within these mixtures is primarily governed by the properties of their surface tension and reaction kinetics when the interactions causing the phase separation depend on concentrations of chemicals that participate in the catalytic activity. We observe self-propelling clusters in a large part of the state-diagram, due to the emergence of multiple mechanisms for self-propulsion, which depend on the interplay between the non-local nature of the chemical interactions, the sizes of the droplets, and thermal fluctuations. 
A prominent emerging feature in this system is that the effective coupling constants controlling the non-reciprocal interactions between two droplets depend inversely on the the droplet sizes, implying that two droplets can cross the threshold of mutually attractive to chasing interaction simply by ripening. The resulting strong and non-trivial dependence of the properties of self-propulsion on the sizes of the droplets could be of significant help in experimental validation of the aforementioned findings. The self-propulsion induced by chemical activity leads to enhanced diffusion of the droplet clusters, providing a method to probe the chemical activity of cellular sub-compartments \emph{in vivo}. We therefore argue our results can help design and analyze chemically active matter, in addition to shedding light on the interplay between activity and non-locality in extended systems. 
We note that a direct estimate for the expected magnitude of the droplet velocity will be difficult to surmise in the absence of quantitative information about the values for the field mobility coefficients $M_a$. However, we can argue that the mechanism that underlies the propulsion is inherently related to the process of droplet coarsening (see the discussion in SM \cite{SuppMat}), which suggests that the propulsion should be visible in the regimes when the coarsening can be experimentally probed. Moreover, the scale of propulsion is directly controlled by the rate of chemical activity $r_b$, which implies that experimental systems with relatively high catalytic rate of reaction such as those involving fast enzymes which produce proton gradients \cite{Testa2021} will be the most promising candidates for detecting the proposed phenomenology. In the future, it would be interesting to extend this study to chemically active suspensions with a large number of components \cite{Parkavousi2025}, incorporate systems with fluid flow to address Marangoni-driven dynamics \cite{maass2016swimming,michelin_self-propulsion_2023}, investigate the cases when metabolic fluxes are themselves the drivers of non-equilibrium phase separation \cite{Cotton2022} and explore possible roles of the emerging phenomenology in origins of life scenarios \cite{OuazanReboul2023b}.

\acknowledgements
We acknowledge support from the Max Planck School Matter to Life and the MaxSynBio Consortium which are jointly funded by the Federal Ministry of Education and Research (BMBF) of Germany and the Max Planck Society.

\bibliography{new_bib,Golestanian,bibliography}
\clearpage
\onecolumngrid
\begin{center}
    \large{\bf End Matter}
\end{center}
\twocolumngrid
\appendix

\section{The role of inter-species interaction}\label{appendix 1}
To test the robustness of our results with respect to the introduction of inter-species interactions, in particular, the existence of the purely attractive chasing states, we add a Flory-Huggins interaction term of the form $\chi \phi_1 \phi_2$ to the free energy.

\begin{figure}[t]
    \centering
    \includegraphics[width=\columnwidth]{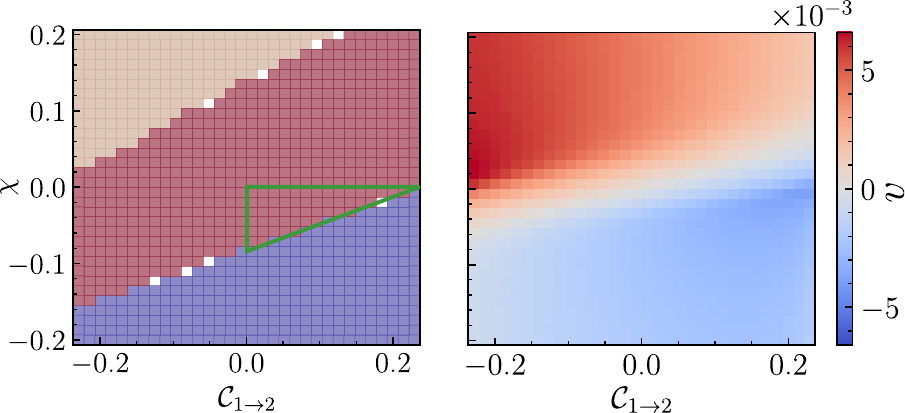}
    \caption{Simulations with non-zero $\chi$. 
    To the left: state diagram categorizing the behaviour of the system, colour as in Fig.~\ref{fig:phase}.
    The green triangle indicates states where all interactions are attractive, but the system still self-propels. 
    To the right: the corresponding steady-state velocity.
    Parameters: $R_1 = 1$, $R_2 = 1.5$ ($\mathcal{R} = \frac{2}{3}$), and $\partial_c \sigma_2(0)= 0.8$.
    }
    \label{fig:chi}
\end{figure}

In Fig.~\ref{fig:chi}, we illustrate the effect of $\chi$ on the state diagram, as obtained from simulation of Eq.~\eqref{eq:eom} in 1D, together with the self-propulsion velocity.
We indeed observe the presence of chasing states at non-zero $\chi$ and positive $\ch{a}{b}$. In particular, when $\chi<0$, the passive interaction between the droplets due to the cross term is also attractive. Nevertheless, simulations show the formation of self-propelling, overlapping states.

\section{Details on numerical simulations}
\label{appendix 2}

For the numerical integration of the equations of motion, Eq.~\eqref{eq:eom}, we have to make a specific choice for free energy. We consider
\begin{align}
   \Fr_a &= \int d^d\bm r\left[
        \frac{1}{2}K(\bm \nabla\phi_a)^2
        + \frac{1}{4} A_a(c)\left(\phi_a^2 - \bar{\phi}_a^2\right)^2
    \right], \\
    A_a &= 
    \left( 
        \sqrt{ A_0 } + { \frac{3}{2 \sqrt{ 2 }} } \alpha_a c 
    \right)^2,
\end{align}
which yields $A_a(c = 0) = A_0$ and $ \partial_c \sigma_a(0) = \alpha_a$.
This particular choice of the Ginzburg-Landau type potential and function $A_a(c)$ is for convenience, but different choices would lead to the same phenomenology (so long as the system is in the linear response regime for the dependence of the surface tension in the chemical concentration).
In simulations, we used $K = 1$, $M_a =1$, $D_c =1$, $-r_1 = r_2 =1$, and $A_0 = 256$.
In the simulations for the phase diagrams, velocity, and position, we include a finite but negligible temperature $T = 10^{-5}$ for improved numerical stability. We furthermore set $c(\bm q=0) = 0$, as this improves stability without affecting the results.

To classify the phases, we initialize two simulations in the configurations (I) and (III), and check if the final configurations are (I), (II), or (III). This gives us data-points such as $[\I,\III]\rightarrow [\I,\I]$, which indicate that both simulations evolve to the (I) configuration. This is thus classified as repulsive. The complete classification rules are as follows: $[\I,\III]\xrightarrow[]{\text{repulsive}} [\I,\I]$, $[\I,\III]\xrightarrow[]{\text{engulfed}} [\III,\III]$, $[\I,\III]\xrightarrow[]{\text{meta-stable}} [\I, \III]$, and $[\I,\III]\xrightarrow[]{\text{overlapping}} [\II, \II]$.

We simulate Eq.~\eqref{eq:eom} using pseudo-spectral methods with periodic boundary conditions,
The system is stiff due to the fourth-order derivative and sharp gradients, which is dealt with by using exponential time-stepping~\cite{coxExponentialTimeDifferencing2002}, as detailed in~\cite{johnsrudEfficientPseudospectralAlgorithms2025}.
For the simulations in $1D$, we use a length $L = 16$ and discretize it in $N = 512$ grid points, such that the thickness of the interface is $\xi = \sqrt{ 2 / A_0 } = \sqrt{ 8 } \dd x$.
In $2D$, we set $N=64$ and $L = 4$ for the simulation of the attractive cases, and $N=128$ and $L = 8$ for the overlapping cases.

\section{Surface-averaged concentrations in different dimensions} \label{app:d-dim-phoresis}

Here we present the  details of the derivation of the expression for effective phoretic velocity of droplets in arbitrary dimensions. The deterministic part of Eq.~\eqref{eq:x-dyn-eq} reads $\dot \x_a=- \frac{d^2 M_a }{R_a \Delta \brp_{a}^2}\frac{\partial\sigma_a}{\partial c}  {\bm \nabla} {\bar c}_{{\rm s},a}(\x_a)$ where ${\bar c}_{{\rm s},a}(\x_a)=\frac{1}{d V_d}\int d\n_a c(\x_a+R_a\n_a)$
We can solve for the chemical concentration by using Eq.~\eqref{eq:chemical} in Fourier space, which yields $c(\x)=\sum_{b}\left(\frac{r_b}{D_c}\right) \int \frac{d^d \q}{(2 \pi)^d} \,\frac{\phi_b(\q)}{q^2+\kappa^2}\,e^{i \q \cdot (\x-\x_b)}$ where $\phi_b(\q)=\int d^d \y \phi_b(\y) e^{-i \q \cdot \y}=\int d\n_b \int_{0}^{R_b} r^{d-1} d r \,e^{-i \q \cdot \n_b r}$ since $\phi_b(\y)=1$ for $|\y| \leq R_b$ and $\phi_b(\y)=0$ for $|\y| > R_b$ in the thin interface limit. Combining these expressions, we obtain
\begin{eqnarray}
&&{\bar c}_{{\rm s},a}(\x_a)=\sum_{b}\left(\frac{r_b}{D_c}\right)\frac{1}{d V_d}\int d\n_a \int d\n_b \int_{0}^{R_b} r^{d-1} d r \nonumber \\
&& \hskip1.3cm \times\int \frac{d^d \q}{(2 \pi)^d} \,\frac{e^{i \q \cdot (\x_a-\x_b+\n_a R_a-\n_b r)}}{q^2+\kappa^2}.\label{eq:cave-fourier-1}  
\end{eqnarray}
By introducing the integral ${\cal J}_d(p)=\frac{\int_0^\pi d\theta \, \sin^{d-2}\theta \, e^{i p \cos\theta}}{\int_0^\pi d\theta\,\sin^{d-2}\theta}$ which can be evaluated as ${\cal J}_d(p)=2^{\frac{d}{2}-1} \Gamma\left(\frac{d}{2}\right) p^{-\frac{d}{2}} \left[d J_{\frac{d}{2}}(p)+p J_{\frac{d}{2}+1}(p)\right]$ in terms of the Bessel function $J_\nu$, we can write a closed form expression for the surface averaged concentration around droplet $a$ as follows
\begin{eqnarray}
&&{\bar c}_{{\rm s},a}(\x_a)=\sum_{b \neq a}\left(\frac{r_b}{D_c}\right)\frac{(d V_d)^2}{(2 \pi)^d}\int_0^{\infty} \frac{d q}{q} \,\frac{{\cal J}_d(q R_a)}{q^2+\kappa^2}  \nonumber \\
&& \hskip1.3cm\times \,{\cal J}_d(q|\x_a-\x_b|) \int_{0}^{q R_b} u^{d-1} d u {\cal J}_d(u),\label{eq:cave-fourier-3}  
\end{eqnarray}
in any dimension $d$. The specific forms ${\cal J}_1(p)=\cos p$, ${\cal J}_2(p)=J_0(p)$, and ${\cal J}_3(p)=\frac{\sin p}{p}$ can be used to write explicit expressions for ${\bar c}_{{\rm s},a}(\x_a)$ in different dimensions, namely
\begin{eqnarray}
&&{\bar c}_{{\rm s},a}^{(d=1)}(x_a)=\frac{2}{\pi}\sum_{b \neq a}\left(\frac{r_b}{D_c}\right)\int_0^{\infty} \frac{d q}{q} \,\frac{\cos (q R_a) \sin (q R_b)}{q^2+\kappa^2}  \nonumber \\
&& \hskip2cm\times \,\cos(q |x_a-x_b|),\label{eq:cave-fourier-d1}  
\end{eqnarray}
for $d=1$, 
\begin{eqnarray}
&&{\bar c}_{{\rm s},a}^{(d=2)}(\x_a)=\sum_{b \neq a}\left(\frac{r_b R_b}{D_c}\right)\int_0^{\infty} d q \,\frac{J_0(q R_a) J_1(q R_b)}{q^2+\kappa^2}  \nonumber \\
&& \hskip2cm\times \,J_0(q |\x_a-\x_b|),\label{eq:cave-fourier-d2}  
\end{eqnarray}
for $d=2$, and 
\begin{eqnarray}
&&{\bar c}_{{\rm s},a}^{(d=3)}(\x_a)=\frac{2}{\pi} \sum_{b \neq a} \frac{1}{|\x_a-\x_b|} \left(\frac{r_b}{D_c R_a}\right)\int_0^{\infty} \frac{d q}{q^3} \,\frac{\sin(q R_a)}{q^2+\kappa^2}  \nonumber \\
&& \hskip0.7cm\times \, \left[\sin (q R_b)-(q R_b)\cos(q R_b)\right]\sin(q |\x_a-\x_b|),\label{eq:cave-fourier-d3}  
\end{eqnarray}
for $d=3$. These expressions are in agreement with those obtained using a more direct integration method.

In the special case where the droplets do not overlap, we obtain a closed expression in the form of generalized phoresis. To this end, we use a formal representation of the Taylor expansion as 
$ {\bar c}_{{\rm s},a}(\x_a) = \frac{1}{d V_d} \smash{\int} d\n_a  e^{R_a\n_a \cdot \bm \nabla} c(\x_a)$ which can be written formally as ${\bar c}_{{\rm s},a}(\x_a)={\cal I}_d\left(R_a \nabla\right) c(\x_a)$
where ${\cal I}_d (b)\equiv {\cal J}_d(p=-i b)$ is given as ${\cal I}_d (b)=2^{\frac{d}{2}-1} \Gamma\left(\frac{d}{2}\right) b^{-\frac{d}{2}} \left[d I_{\frac{d}{2}}(b)+b I_{\frac{d}{2}+1}(b)\right]$ in terms of the modified Bessel function of the first kind $I_\nu$. In particular, we have ${\cal I}_1(b)=\cosh b$, ${\cal I}_2(b)=I_0(b)$, and ${\cal I}_3(b)=\frac{\sinh b}{b}$. Since we are interested in contributions to $c$ from $b\neq a$ and because the droplets do not overlap, the right-hand (source) side of Eq.~\eqref{eq:chemical} is zero in the region of interest, and we can write $\nabla^{2n} c =\kappa^{2 n} c$ for $n \geq 1$. Combining these results, we obtain $\dot \x_a =- \frac{d^2 M_a }{R_a \Delta \brp_{a}^2}\frac{\partial\sigma_a}{\partial c} \,{\cal I}_d\left(\kappa R_a\right) {\bm \nabla} c(\x_a)$.
Note that the effective phoretic mobility arising from this mechanism shown in Eq.~\eqref{eq:phoresis-result-d} exhibits a leading order $1/R$ size dependence in any dimension $d$, which is in sharp contrast with the case of standard diffusiophoresis where the mobility is independent of size \cite{golestanianPhoreticActiveMatter2022}. Moreover, in the derivation of this result no thin slip-layer assumption is used \cite{anderson1989,AgudoCanalejo2018}, and near-field effects are taken into consideration \cite{Nasouri2020a,Nasouri2020}, as represented by the function ${\cal I}_d\left(\kappa R_a\right)$, which introduces additional $d$-dependent size dependence.

\vspace{3.5mm}
\section{Effective potentials in 1D}
\label{app:1D}

We report here the expression for $W_{\pm}$ in Eq.\eqref{eq:dynLangevin} for $d=1$.
We consider the three subdomains corresponding to the three possible configurations, as illustrated in Fig.~\ref{fig:possdroplets} (c):
(I) disjoint, (II) partially overlapping, and (III) completely overlapping.  These correspond to $|x|>R_1 + R_2$, $R_2-R_1<|x|<R_1+R_2$, and $|x|<R_2 - R_1$ respectively. 
The effective potentials $W_{\pm}(x)$ are given by:
\begin{widetext}
\begin{equation}
    \label{eq:effpot}
   W_{\pm}(x)= \frac{1}{\mathcal{M}}\begin{cases}
    \vspace{0.1cm}
        - 2^{-\frac{(1\pm1)}{2}}e^{-|x|}\left[\frac{\chto}{R_2}\cosh R_2\sinh R_1\mp\frac{\chot}{R_1}\cosh R_1\sinh R_2\right]+\text{const}, 
        &\text{(I)\phantom{II}}
        \\
    \vspace{0.1cm}
        2^{-\frac{(3\pm1)}{2}}e^{- (R_1+R_2)}\left[\left(\frac{\chto}{R_2}\pm\frac{\chot}{R_1} \right)\cosh x+e^{-  |x|}\left(\frac{\chto}{2R_2}\mp\frac{\chot}{2R_1}\right)(e^{2R_2}-e^{2R_1})\right], 
        & 
        \text{(II)\phantom{I}}
        \\    
       -2^{-\frac{(1\pm1)}{2}}e^{-R_2}\left[\frac{\chto}{ R_2}\sinh R_1\pm\frac{\chot}{ R_1}\cosh  R_1\right]\cosh x+\text{const}. 
       & 
       \text{(III)}
    \end{cases}
\end{equation}
\end{widetext}

\onecolumngrid
\clearpage

\begin{center}
\textbf{\large Non-reciprocal interactions between condensates in chemically active mixtures} \\ \textit{\large Supplemental Material}\\[1em]
Jacopo Romano$^{1,2}$, Martin Kjøllesdal Johnsrud$^{1}$, Beno\^it Mahault$^{1}$ and Ramin Golestanian$^{1,3}$\\[0.2em]
 {\it$^{1}$~Max Planck Institute for Dynamics and Self-Organization (MPI-DS), 37077 G\"ottingen, Germany.}\\
{\it $^{2}$~SISSA --- International School for Advanced Studies and INFN, via Bonomea 265, 34136 Trieste, Italy}\\
{\it $^{3}$~Rudolf Peierls Centre for Theoretical Physics, University of Oxford, Oxford OX1 3PU, United Kingdom.}
\end{center}
\vspace{1cm}
\setcounter{equation}{0}
\setcounter{figure}{0}
\setcounter{section}{0}
\renewcommand{\theequation}{S\arabic{equation}}
\renewcommand{\thefigure}{S\arabic{figure}}
\renewcommand{\thesection}{S.\Roman{section}}
\subsection{Rescaling of Eq.~\eqref{eq:eom} in the linear chemical response regime}

The force acting on a droplet interface in response to chemical modulations depends on how the surface tension $\sigma$ and the binodal densities $\bar \phi_\pm$ vary with the chemical concentration $c(\x)$. For weak coupling, one can linearize these dependencies as: $\sigma_a(c)=\sigma_{a}(0)+\partial_c\sigma_{a}(0)c$, $\phi_{a,\pm}(c)=\bar\phi_{a,\pm}(0)+\partial_c\bar\phi_{a,\pm}(0) c$, where the subscript $a$ denote the droplet species. The velocity $v_a$ of a droplet of species $a$ is then a function of the response coefficients:
$v_a = v_a(\partial_c \sigma_a(0), \partial_c\bar{\phi}_{a,\pm}(0),c)$, with $v_a=0$ for $c={\rm const}$. 
In the weak-response regime, we retain only terms up to first order in the responses.

We argue now that the dependence of $v_a$ on $\partial_c \bar{\phi}_{a,\pm}(0)$ is negligible at first order. In fact, using the results of Ref.~\cite{romano2025dynamics} (see in particular Sec.~4), we note that this dependence arises from two distinct contributions. The first is a correction of order $\mathcal{O}(\partial_c \bar{\phi}_{a,\pm}(0))$ to the droplet mobility.
Since the force generated by chemical gradients is of order $\mathcal{O}(\partial_c \sigma_a(0))$, this correction affects the velocity only at the subdominant order $\mathcal{O}(\partial_c\bar{\phi}_{a,\pm}(0)\partial_c \sigma_a(0))$. The second contribution corresponds to an effective interfacial force that originates from mass displacements associated with changes in the binodal densities. This term scales as $\mathcal{O}(\partial_c\bar{\phi}_{a,\pm}(0)\,v_b)$, where $v_b$ is the velocity of the droplet of the other species. However, since $v_b = \mathcal{O}(\partial_c \sigma_b(0))$, this contribution is also subdominant. Therefore, in our analysis of the droplet dynamics in one dimension and for weak responses, we can assume $\partial_c\bar{\phi}_{a,\pm}(0) = 0$ without loss of generality.
In higher dimensions, the analysis of systems with space-dependent binodal densities becomes more challenging, since for interfaces with finite curvature the dependence of $v_a$ on $\partial_c\bar{\phi}_{a,\pm}(0)$ is no longer subdominant. For these systems, we nevertheless set $\partial_c \bar{\phi}_{a,\pm}(0) = 0$ to simplify the analysis, although the generalization to include this correction is not expected to change the qualitative features of the results.

We now discuss the transformations leading to the reduction of the parameter space presented in the main text. 
Since $\bar{\phi}_{\pm,a}$ are constants we can rescale the fields as
\begin{equation}
    \label{eq:fieldtrans}
    \phi_a\rightarrow \brp_{a,-}+\Dbp_a\,\phi_a,\hspace{0.5cm}c \rightarrow \frac{r_1\brp_{1,-}+r_2\brp_{2,-}}{\gamma}+c,
\end{equation}
such that the newly defined density fields for the species go from $0$ outside of the droplet to $1$ inside of it, while the chemical field has been rescaled by its background value.
The equation for the chemical in \eqref{eq:LRNRCH} naturally introduces a timescale $t_{\gamma}=1/\gamma$ and a length scale $\kappa^{-1}=\sqrt{D_c/\gamma}$. It is then natural to rescale \eqref{eq:LRNRCH} with respect to these two, so that both length and time become dimensionless. The free energy has a natural unit of energy ${\cal E}$ related to the energy scales of the molecular interactions, which we also use to  non-dimensionalize the temperature. These transformations account for rescaling the coefficients of \eqref{eq:LRNRCH} to:
\begin{gather}
\label{eq:coefftrans}
    r\rightarrow \sqrt{\frac{D_c}{\gamma}} r,\hspace{0.5cm}t\rightarrow \frac{t}{\gamma},\hspace{0.5cm} r_a\rightarrow\frac{\gamma}{\Dbp_a}r_a,\hspace{0.5cm} M_a\rightarrow\frac{M_a (\Dbp_a)^2 D_c^{1+d/2}}{{\cal E}\gamma^{d/2}},\hspace{0.5cm}\Fr \rightarrow {\cal E} \Fr,\hspace{0.5cm}T \rightarrow {\cal E} T.
\end{gather}

\subsection{Ripening in multi-droplet systems}

In the main text, we focused on systems composed of a single isolated droplet per species. However, when multiple droplets of the same species coexist in the mixture, Ostwald ripening can cause their sizes to evolve on timescales comparable to those of droplet motion. To illustrate this effect, we consider the simple but instructive case of two droplets of species $a$, separated by a distance $L$ in $d$ spatial dimensions. We neglect shape deformations and assume that both droplets remain approximately spherical throughout their evolution.

The interface point $\sX$ on droplet $I=1,2$ move with velocity
$
\dot{\sX} = \dot R_I\, \mathbf{n}(\sX) + \dot{\mathbf{x}}_I,
$
where $R_I$ and $\mathbf{x}_I$ denote the droplet radius and centre of mass, respectively, and $\mathbf{n}(\sX)$ is the outward normal unit vector at $\sX$. We denote by $\mathcal{S}_I$ and $\mathcal{V}_I$ the surface area and volume of droplet $I$. Furthermore, we define the surface-averaged force on droplet $I$ as
\begin{equation}
    F_I = \frac{1}{\mathcal{S}_I} \int_{{S}_I} dS_{\sX} \left[ \frac{d-1}{R_I}\, \sigma(c(\sX)) + \mathbf{n}(\sX)\cdot\nabla\sigma(c(\sX)) \right].
\end{equation}
Applying $\frac{1}{\mathcal{S}_I}\int_{{S}_I} dS_{\sX}$ to the interface equations of Ref.~\cite{romano2025dynamics} for $I=1,2$, we obtain
\begin{equation}
\label{eqapp:ripeningcases}
\begin{cases}
    \dot R_1\, \mathcal{S}_1\, G_a(R_1) + \dot R_2\, \mathcal{S}_2\, G_a(L) + \dot{\mathbf{x}}_2 \cdot \nabla G_a(L)\, \mathcal{V}_2 = F_1 - \mu, \\[4pt]
    \dot R_1\, \mathcal{S}_1\, G_a(R_1) + \dot R_1\, \mathcal{S}_1\, G_a(L) + \dot{\mathbf{x}}_1 \cdot \nabla G_a(L)\, \mathcal{V}_1 = F_2 - \mu, \\[4pt]
    \dot R_1\, \mathcal{S}_1 + \dot R_2\, \mathcal{S}_2 = 0,
\end{cases}
\end{equation}
where $G_a$ is the Green’s function of the Laplace equation, satisfying $M_a \nabla^2 G_a(\mathbf{x}) = \delta^d(\mathbf{x})$.

We are interested in the regime of well-separated droplets, $L \gg R_I$. In this limit,
$
\dot{\mathbf{x}}_I \cdot \nabla G_a(L)\, \mathcal{V}_I \ll \dot R_I\, \mathcal{S}_I\, G_a(L),
$
so the centre-of-mass motion has a negligible influence on ripening. Solving Eq.~\eqref{eqapp:ripeningcases} gives the ripening rate:
\begin{equation}
\label{eqapp:ripeningeq}
    \frac{d\mathcal{V}_1}{dt} = -\frac{d\mathcal{V}_2}{dt}
    = \frac{F_2 - F_1}{\,2 G_a(L) - G_a(R_2) - G_a(R_1)\,}.
\end{equation}

Equation~\eqref{eqapp:ripeningeq} shows that the impact of ripening on droplet dynamics depends strongly on dimensionality. For $d \leq 2$, the Green’s function $G_a(L)$ diverges in magnitude as $L$ increases, implying that droplets exchange only negligible amounts of mass once they are sufficiently far apart. Consequently, for $d \leq 2$, clusters of droplets can persist and travel distances much larger than their size, provided the droplets of species $a$ and $b$ start at separations comparable to their radii at $t=0$.

In contrast, for $d > 2$, one has $G_a(L) \to 0$ as $L \to \infty$, and the ripening speed saturates at a value comparable to the centre-of-mass speed. In this regime, ripening takes place on timescales comparable to cluster formation and motion. Therefore, the dynamics described in the main text will predominantly occur for relatively larger droplets interacting with other species, as they grow at the expense of smaller droplets of their own species in their local neighbourhoods.

\subsection{Droplet interaction in 3D using direct integration}
\label{app:dropinteraction}

We provide a direct calculation of the droplet interaction in $d=3$ and use it to determine the phase transition conditions for the diagram in Fig. \ref{fig:pasedia3ddecay}(a). Without loss of generality (for $T=0$), we assume that the droplets are aligned along the vertical axis, such that the motion is restricted to the $z$-direction. Using Eq. \eqref{eq:x-dyn-eq} we obtain
\begin{equation}
    \label{eq:zCoMmotion}
    \dot z_a=-\dfrac{9M_a}{4\pi R_a}\frac{\partial\sigma_a}{\partial c} \int d\Omega \cos\theta \,\partial_z c_b(\theta,\varphi),
\end{equation}
since only the chemical produced by droplet $b\neq a$ affects the velocity of droplet $a$.
Droplet $b$ induces a perturbation $c_b$ of the chemical field, with respect to the background value given as
\begin{equation}
   \label{eqapp:chemfielddecay3d}
   c_b(\bm r)=\begin{cases}
            c^{\text{in}}_b(\bm r)=1 -e^{-R_b}(1 + R_b) \, \dfrac{\sinh |\bm r-\x_b|}{|\bm r-\x_b|}&\text{ for }\;\; |\bm r-\x_b|<R_b,\vspace{0.3cm}\\
            c^{\text{out}}_b(\bm r)=(R_b \cosh R_b-\sinh R_b)\,\dfrac{e^{-|\bm r-\x_b|}}{|\bm r-\x_b|}& \text{ for }\;\; |\bm r-\x_b|>R_b.
            \end{cases}
\end{equation}
\begin{figure}[!tb]
\centering\includegraphics[width=.8\textwidth]{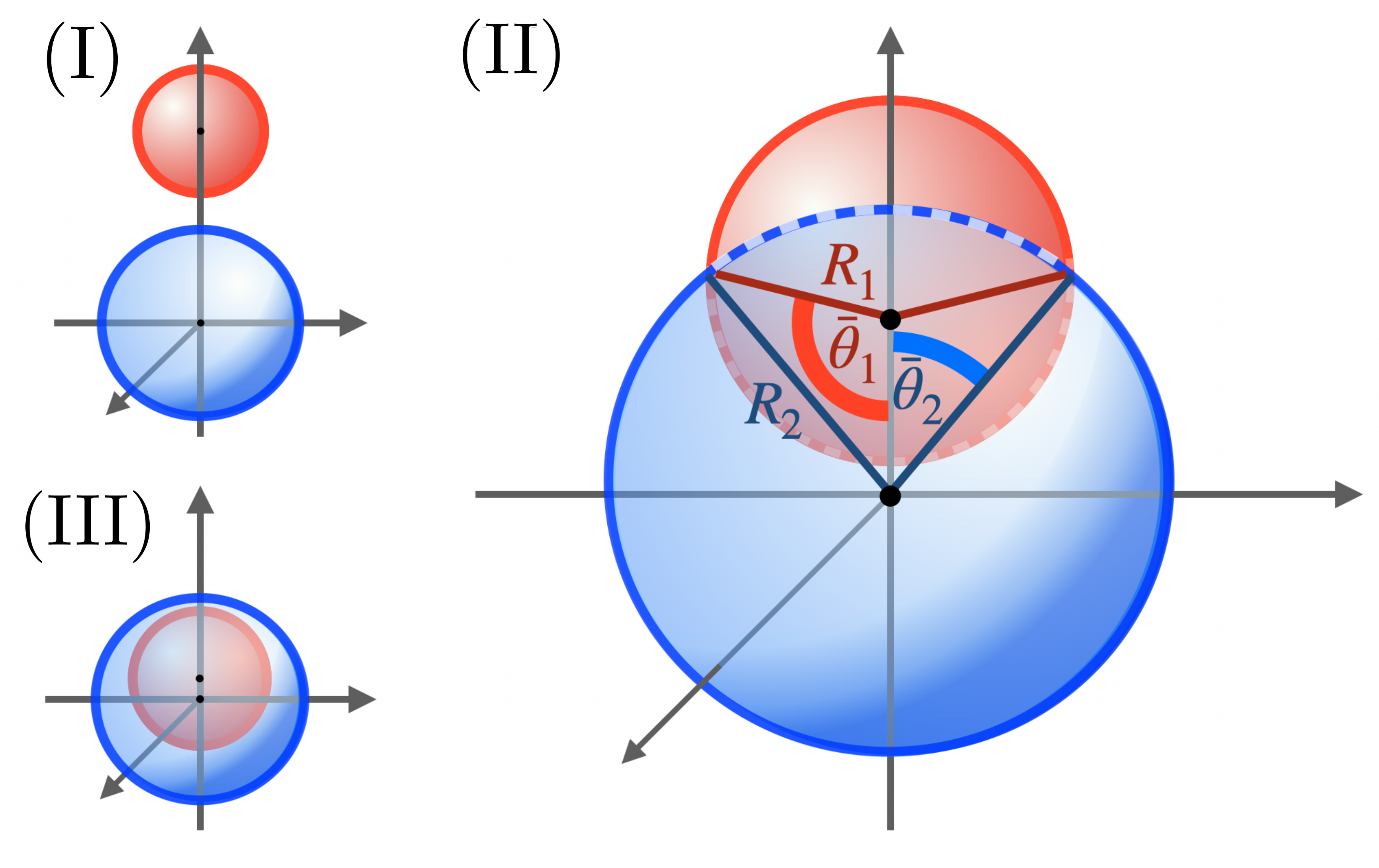}
    \caption{
    Schematic of the three different droplet configurations, (I), (II) and (III), for $a=1$ in blue and $b=2$ in red.
    }
    \label{fig:3dCalc}
\end{figure}

The most difficult sub-case is the one in which the droplets intersect, corresponding to configuration (II). We place droplet $a$ at the origin and droplet $b$ above it. The droplets partially intersect such that a portion of the interface of $a$ lies inside droplet $b$. We denote by $\bar\Omega_a$ the solid angle subtended by the part of droplet $a$ that is immersed in $b$, and by $\bar\theta_a$ the corresponding polar angle, as illustrated in Fig.~\ref{fig:3dCalc}(b).

Let $|z| = |z_b - z_a|$ be the centre-to-centre distance. The integral in Eq.~\eqref{eq:zCoMmotion} can be evaluated piecewise, using the ``inside'' expression for $c_b$ (Eq.~\eqref{eqapp:chemfielddecay3d}) for $0 \le \theta \le \bar\theta_a$, and the ``outside'' expression for $\bar\theta_a \le \theta \le \pi$. The intersection angle is given by the geometric relation
\begin{equation}
    \label{eq:int_angle}
    \bt_a=\frac{z^2+R_a^2-R_b^2}{2|z| R_a},
\end{equation}
The equations of motion follow from evaluating the integrals in Eq.~\eqref{eq:zCoMmotion} for each droplet and substituting Eq.~\eqref{eq:int_angle} to express the forces in terms of the relative distance $z$. We obtain
\begin{equation}
    \label{eq:disjointintegral}
    \dot z_a = - \dfrac{9M_a \partial_c \sigma_a(0)}{2 R_a}
    \left[\int_{\bt_a}^{\pi}d\theta\,\sin\theta \cos\theta \,\partial_zc^\text{in}_b(R_a,\theta) +
    \int^{\bt_a}_{-\pi}d\theta\,\sin\theta \cos\theta \,\partial_zc^\text{out}_b(R_a,\theta)
    \right].
\end{equation}
The (I) and (III) subcases can be obtained from \eqref{eq:disjointintegral} by setting $\bt_a=-\pi$ for the smaller droplet in configuration (II), and $\bt_a=\pi$ in all the other cases.  We performed integral in \eqref{eq:disjointintegral} analytically using the symbolic manipulation software \emph{Mathematica} and use it to study the effective potential in \eqref{eq:dynpotential} for the 3D case. In configuration (III), we find that the fixed point at $x=0$ is a minimum if the following condition holds
\begin{equation}
    \label{eq:stabnest3ddecay}
    \chto<\dfrac{(R_2+1)R_2^2\sinh R_1 }{R_1^2[R_1 \cosh R_1-\sinh R_1]}\chot,
\end{equation}
leading to (meta-)stable engulfed droplets. The condition for repulsive droplets in configuration (I) is
\begin{equation}
    \label{eq:repulsive3ddecay}
    \chto<-\dfrac{R_2^2 \sinh R_1 [R_2 \cosh R_2 - \sinh R_2 ]}{R_1^2 \sinh R_2 [R_1 \cosh R_1 - \sinh R_1]}\chot,
\end{equation}
whereas for
\begin{align}\label{eq:stabclu3ddecay}
\chto>\dfrac{(R_2+1)R_2^2\sinh R_1 }{R_1^2[R_1 \cosh R_1 - \sinh R_1]}\chot
    && \text{and} &&
    \chto>-\dfrac{ R_2^2 \sinh R_1 [R_2 \cosh R_2 - \sinh R_2 ]}{R_1^2 \sinh R_2 [R_1 \cosh R_1 - \sinh R_1]}\chot,
    \end{align}
$V$ is a symmetry-breaking Mexican hat potential leading to the formation of self-propelling clusters in configuration (II). We use the conditions \eqref{eq:stabclu3ddecay}, \eqref{eq:repulsive3ddecay}, \eqref{eq:stabclu3ddecay} to build the phase diagram of the model, leading to the results in Fig.~\ref{fig:pasedia3ddecay} of the main text.

\subsection{Stochastic dynamics $d$-dimensional droplets}

We now derive the expression for the stochastic force acting on the droplets centres of mass at finite temperature ($T \neq 0$) in the relevant case $\bar{\phi}_{a,+} = 1$ and $\bar{\phi}_{a,-} = 0$, although the analysis can be readily generalized. The first step is to determine the effective noise acting on a point of the interface of each droplet due to the stochastic current 
$\bm{J}_{a}^s = \sqrt{2 M_a T}\,\boldsymbol{\nabla} \cdot \boldsymbol{\Lambda}_a$. 
This problem has been studied in the literature for both passive and active systems~\cite{Cates2023nucleation}.
Our derivation is based on a generalization of the results in Ref.~\cite{romano2025dynamics}. 

We start by inverting the Laplacian in Eq.~\eqref{eq:LRNRCH} to obtain
\begin{equation}
    \int d^d {\bm r}' G_a(\bm r-\bm r')\,\partial_t\phi_a=\elD{\Fr_a(\phi_a,c)}{\phi_a}(\bm r,t)+\bar\mu^s_a(\bm r,t)-\bar\mu_{\infty,a},
\end{equation}
where $G_a(\bm r)$ is the Green’s function of the $M_a\nabla^2$ operator 
[i.e., the solution of $M_a \nabla^2 G_a(\bm r) = \delta^d(\bm r)$], and
\begin{equation}
    \label{eqapp:stochpot}
    \bar\mu^s_a(\bm r,t)=\int d^d {\bm r}' G_a(\bm r-\bm r')\bm\nabla_{\bm r'} \cdot\lt[\sqrt{2M_a}\Lam_a(\bm r',t)\rt].
\end{equation}
Comparing Eq.~\eqref{eqapp:stochpot} with the results of Ref.~\cite{romano2025dynamics} 
(see in particular Sec.~2.2), we find that $\bar{\mu}^s_a$ enters as a stochastic thermodynamic force (denoted $\bar{\mu}(\bm r, t)$ in Ref.~\cite{romano2025dynamics}) in the interface equations. We calculate its variance using Eq.~\eqref{eqapp:stochpot} and find
\begin{equation}
    \label{eqapp:noisecorr}
\langle\bar\mu^s_a(\bm r,t)\bar\mu^s_b(\bm r',t')\rangle=-2 T\delta_{ab}G_a(\bm r-\bm r')\delta(t-t').
\end{equation}
The noise acting on the centre of a droplet, $\x_a$, can then be derived from the noise on its interface. In 1D, this corresponds to the following system of equations
\begin{align}
    \label{eqapp:simplemobnoise}
    \begin{cases}
         -\dot x_a \, G_a(2 R_I)={\partial_x \sigma}(x_{a,r})+\bar\mu^s_a(x_{a,r},t)-\bar\mu_{a,\infty}, \\
         \dot x_a\, G_a(2 R_a)  =-{\partial_x \sigma}(x_{a,l})+\mu^r_I(x_{a,l},t)-\bar\mu_{a,\infty}.
    \end{cases}
\end{align}
Solving \eqref{eqapp:simplemobnoise} for $\dot x_a$ we find the following equation of motion for the droplet at non-zero temperature
\begin{align}
    \frac{2 R_a}{M_a}\,\dot x_a=-[{\partial_x \sigma}(x_{a,r})+{\partial_x \sigma}(x_{a,l})]+2\sqrt{\frac{R_a}{M_a}T}\,\eta_a(t), 
\end{align}
where $\eta_a(t)$ is a Gaussian white noise as defined in the main text.

In higher dimensions, the expression of for the speed of the droplet's centre is a generalization of the results obtained for the determinist case in Ref.~\cite{romano2025dynamics}. In particular, we find
\begin{equation}
\label{eqapp:stoch3d}
     \frac{\pazocal{V}_a}{d M_a}\dot \x_a = -\int_{S_a}dS_{\sX} \; \nabla\sigma_a(\sX)+\int_{S_a}dS_{\sX} \; \n(\sX)\mu^s_a(\sX,t),
\end{equation}
where $\int_{S_a}dS_{\sX}$ denotes a surface integral over the droplet interface in which ${\sX}$ is the integration variable. The first integral in the r.h.s. of Eq. \eqref{eqapp:stoch3d} gives the deterministic force already discussed for the 3D case in Section \ref{app:dropinteraction}. The second integral gives the stochastic force on the droplet's centre. Using \eqref{eqapp:noisecorr} we find that its correlator is given by
\begin{equation}
    \label{eq:highdimnoise}
    \int_{S_a}dS_{\sX} \int_{S_b} dS_{\sX'}  \; \hat n_i(\sX)\hat n_j(\sX')\langle\mu^s_a(\sX,t)\; \mu^s_b(\sX',t')\rangle=2T\frac{\pazocal{V}_a}{d M_a}\delta_{ij}\delta_{ab}\delta(t-t'),
\end{equation}
leading to the noise correlators reported in the main text.

\subsection{The effect of hydrodynamics}

The description of phase separation in momentum--conserving systems can be formulated using a diffusive phase field and a hydrodynamic flow field $\bm u(\bm x,t)$ describing convective transport. This leads to the so-called Model~H in the classification of Ref.~\cite{HalperinHohenberg1977}. For a single species at low Reynolds number (the generalization to multiple species being straightforward), Model~H reads:
\begin{equation}
\label{eqresp:modelH}
\begin{cases}
\partial_t \phi + \bm u \cdot \nabla \phi
= M \nabla^2 \dfrac{\delta \mathcal{F}[\phi]}{\delta \phi}, \\[10pt]
\eta \nabla^2 \bm u - \nabla P - \phi \nabla \dfrac{\delta \mathcal{F}[\phi]}{\delta \phi} = 0 ,
\end{cases}
\end{equation}
where $M$ is the mobility of the phase field, $\mathcal{F}[\phi]$ is a free--energy functional, $P$ is the pressure enforcing incompressibility: $\nabla \cdot \bm u = 0$, and $\eta$ denotes the dynamic viscosity of the mixture.
One can show from dimensional analysis that diffusive transport dominates over length scales $\ell \lesssim \sqrt{M\eta}$~\cite{Bray2002review}, while at larger scales transport is governed by convection (for example, Marangoni flows in the case of interfacial transport). 

This scaling behaviour can be understood by focusing on the dynamics at the scale of an individual droplet of radius $R$. The hydrodynamic Marangoni-type velocity scale can be obtained by equating the scaling form of the strain rate with the scaling form of the stress, namely, $\eta v_{\rm h}/R \sim \delta \sigma /R \sim \delta c \cdot \frac{\partial \sigma}{\partial c} /R $, which yields $v_{\rm h} \sim \delta c \cdot \frac{\partial \sigma}{\partial c}/\eta$. For the diffusive process, we have $v_{\rm d} \sim [M \frac{\partial \sigma}{\partial c} {\cal I}_d(\kappa R)/R] (\delta c/R) \sim \delta c \cdot \frac{\partial \sigma}{\partial c} M {\cal I}_d(\kappa R)/R^2$. This yields
\begin{equation}
    \frac{v_{\rm h}}{v_{\rm d}} \sim \frac{1}{M \eta } \cdot \frac{R^2}{\mathcal{I}_d(\kappa R)}=\frac{\kappa^{-2}}{M \eta } \cdot \frac{(\kappa R)^2}{\mathcal{I}_d(\kappa R)},
\end{equation}
which implies that in addition to the effect discussed above at small scales, hydrodynamic effects will also be suppressed by the exponential dependence of ${\mathcal{I}_d(\kappa R)}$ at large $R$. For the interaction between different droplets the typical distance between droplets will also enter the expression but it will not change the conclusion. Therefore, we conclude that hydrodynamic effects will not be able to compete with diffusive transport in the majority of circumstances, except for cases when $\kappa R \sim 1$ and $\kappa^{-1} \gtrsim \sqrt{M\eta}$.

In the absence of direct quantitative information about the material parameters of the mixture such as the mobility $M$, one can probe the importance of hydrodynamics by monitoring the coarsening dynamics, as it is known that diffusive transport is characterized by a slow $R \sim t^{1/3}$ growth regime whereas hydrodynamic transport exhibits a linear growth regime, i.e., $R \sim t$ \cite{Bray2002review}. We further note that in quasi-two-dimensional systems such as~\cite{tateno2026diffusive} diffusive transport will always dominate over hydrodynamics regardless of the scale of interest (see Ref.~\cite{romano2025dynamics} for more details).

\end{document}